\documentclass[12pt]{article}
\usepackage{graphicx}
\begin{document}

\title{NONUNIFORM DONNAN EQUILIBRIUM WITHIN BACTERIOPHAGES PACKED WITH DNA}
\author{Theo Odijk$^*$ \\
Section Theory of Complex Fluids \\
Faculty of Applied Sciences Delft \\
University of Technology Delft, the Netherlands \\
\\
Flodder Slok \\
Department of Zoology \\
University of Amsterdam, the Netherlands}

\maketitle

\begin{abstract}
The curvature stress of DNA packed inside a phage is balanced against its
electrostatic self-interaction. The DNA density is supposed nonuniform and
as a result the Donnan effect is also inhomogeneous. The coarse-grained DNA
density is a nonlinear function of the DNA radius of curvature at a given
position inside the bacteriophage. It turns out that a region (or regions)
exists totally free from DNA. The size of such holes is computed.
\end{abstract}

\vskip 40pt
\noindent
$^*$ Address for correspondence : \\
T. Odijk, P.O. Box 11036, 2301 EA Leiden, the Netherlands \\
E-mail: odijktcf@wanadoo.nl

\pagebreak

The packing of DNA in cells and viruses are problems in the
nanothermodynamics \cite{ref1} of polyelectrolytes i.e. the systems are
small and hence the free energies need not be extensive. A case in point is
the compaction of DNA inside bacteriophages where the extensive self-energy
of the DNA spool is balanced against the non-extensive energy of curvature
arising mainly from a region of tightly wound DNA surrounding a hole \cite
{ref2}. The degree of non- extensivity depends on the problem at hand. For
fairly large DNA globules that are compacted in solution, the shape is
determined by non-extensive terms whereas the volume is not \cite{ref3, ref4}.

In previous work \cite{ref2, ref3, ref4} we assumed the DNA density in a
coarse-grained continuum approximation to be homogeneous, but we relax this
supposition here. This introduces a peculiar effect : the Donnan effect is
now also nonuniform and it happens to compete with the curvature stress
which increases sharply in tightly wound regions. Although the uniform
approximation of \cite{ref2} agrees well with the DNA spacings within
bacteriophage T7 determined experimentally \cite{ref5}, the forces
encountered in single molecule experiments \cite{ref6} of DNA being forced
into Phi 29 phage by the connector motor seem to overestimate the predicted
forces \cite{ref2, ref7}. Thus a reinvestigation of at least one assumption
is warranted.

Close-packed DNA in vitro has often been found to be a hexagonal liquid
crystal \cite{ref8, ref9}. Sometimes, for long polydisperse DNA, the
hexagonal phase appears to be precluded in favor of a possible ''hexatic''
phase \cite{ref10} but a full investigation of the algebraic decay of the
ovientational order has yet to be carried out. Here, for the sake of
definiteness, we assume the DNA lattice inside a phage is hexagonal (for
general, biologically oriented reviews on DNA packaging in bacteriophages, 
\cite{ref11, ref12}). However, such a lattice is probably rarely ever ideal
but is bound to contain DNA crossovers leading to knotted entanglements upon
release of the DNA from the phage \cite{ref13, ref14}.

Theories of slightly deformed hexagonal phases have been set up previously 
\cite{ref15, ref16}. They apply to macroscopic or extensive systems. Here,
our compacted DNA is of mesoscopic dimensions and its properties are
dominated by nonextensive curvature stresses so we simply disregard
fluctuations in the director $n(\mathbf{r})$ and coarse-grained DNA density
$\rho (\mathbf{r})$ at position $\mathbf{r}$. The linear double-stranded
DNA has only two ends so to a good approximation a well-known continuity
law applies \cite{ref17, ref18}. 
\begin{equation}
\mathbf{\nabla} \cdot \rho \mathbf{n} = 0  \label{eq1}
\end{equation}
We next assume absence of splay ($\mathbf{\nabla} \cdot \mathbf{n}=0$)
since the splay elastic constant is expected to be very large. In general, at constant
density, we would deal with the theory of so-called developable domains in
describing the deformed hexagonal liquid crystal \cite{ref19}. We here
consider deformations of nonuniform densities in such a way that the
director remains perpendicular to the density gradient. Under these
conditions eq.(\ref{eq1}) is obeyed.

Now the simplest total free energy of the DNA coil as a functional of the
density $\rho (\mathbf{r})$ seems to be 
\begin{equation}
F_{tot} = F_{int}\left[ \rho (\mathbf{r}) \right]
+ U_{b}\left[ \rho (\mathbf{r}) \right]
\label{eq2}
\end{equation}
where $F_{int}$ is the free energy of the DNA interacting with itself in a
hypothetically straightened hexagonal lattice (though it is inhomogeneous
across the plane perpendicular to its central axis) and $U_{b}$ is the
bending energy given by 
\begin{equation}
U_{b} = \frac{1}{2} P \, k_{B} T \int_{0}^{L}ds\ R_{c}^{-2}(s)
\label{eq3}
\end{equation}
Here, $P$ is the DNA persistence length, $k_{B}$ is Boltzmann's constant, $T$
is the temperature and $R_{c}$\ is the radius of curvature of the DNA
wormlike coil at position $s$\ on the DNA contour of total length $L$. As
already stated, undulations of the DNA will be disregarded in the tightly
packed state we focus on. If we define the surface area $S(\mathbf{r})$\ of
the unit cell of the deformed hexagonal lattice at position $\mathbf{r}$, we
may introduce a continuum approximation $S(\mathbf{r})ds=d\mathbf{r}$ and $%
\rho (\mathbf{r})=1/AS(\mathbf{r})$. The length of a DNA nonomer is $A$ so
that $\rho (\mathbf{r})$\ is a monomer density from now on. 
\begin{equation}
U_{b} = \frac{1}{2}P A \, k_{B}T \int_{V} d\mathbf{r} \frac{\rho (\mathbf{r})}
{R_{c}^{2}(\mathbf{r})}
\label{eq4}
\end{equation}
The number of monomers inside the phage of volume $V$\ is kept fixed. 
\begin{equation}
N = \frac{L}{A} = \int_{V}d\mathbf{r} \rho (\mathbf{r)}
\label{eq5}
\end{equation}
Supposing gradients of $\rho (\mathbf{r)}$ may be disregarded in $F_{int}$,
we now minimize the total free energy with respect to $\rho (\mathbf{r)}$\
while accounting for the constraint eq.(\ref{eq5}).
\begin{equation}
\mu (\mathbf{r}) + \frac{P A \, k_{B} T}{2 R_{c}^{2}(\mathbf{r})}
+ const. = 0
\label{eq6}
\end{equation}
We have introduced the inhomogeneous bulk chemical potential of the DNA
monomers ($\mu =\delta F_{int}/\delta \rho (\mathbf{r})$). It is readily
seen that the curvature term becomes appreciable for tightly wound regions
appropriate for DNA packed in phages.

Next, we need an expression for the DNA chemical potential. The phage is in
thermodynamic equilibrium with a large reservoir containing a univalent salt
of concentration $c_{s}$.

The phage is presumed to be free from multivalent ions. We first focus on a
homogeneous hexagonal lattice of DNA and its Donnan equilibrium with such a
reservoir. One convenient approximation for the osmotic pressure is due to
Oosawa \cite{ref20, ref21}. For our purposes, we set the osmotic coefficient
equal to a constant so we write the osmotic pressure $\pi _{os}$\ of the
uniform bulk hexagonal phase with respect to the buffer as 
\begin{eqnarray}
\pi _{os} &=&\left( \frac{A\rho _{0}k_{B}T}{2Q}\right) h(w) \nonumber \\
&=&\left( \frac{k_{B}T}{3^{\frac{1}{2}}H_{0}^{2}Q}\right) h(w)
\label{eq7}
\end{eqnarray}
\[
w\equiv 2.3^{\frac{1}{2}}H_{0}^{2} Q c_{s} = 4 S_{0} Q c_{s} 
\]
\begin{equation}
h(x)\equiv (1+x^{2})^{\frac{1}{2}}-x  \label{eq8}
\end{equation}
The uniform monomer density of the lattice is $\rho _{0}$ and the Bjerrum
length $Q=q^{2}/\varepsilon k_{B}T$\ in terms of the elementary charge $q$
and the permittivity of the solvent $\varepsilon $. For a uniform hexagonal
lattice, the area of the unit cell is $S_{0}=3^{\frac{1}{2}}H_{0}^{2}/2$
where $H_{0}$\ is the lattice spacing i.e. the distance between centeraxes
of adjacent DNA helices ($\rho _{0}=1/AS_{0}$). Eq.(\ref{eq7}) is, of course, an
approximation though, on the whole, it does agree with recent careful
measurements of the osmotic pressure of DNA suspensions by Raspaud et al 
\cite{ref22} (to a good approximation, the pressure conforms to a universal
form $\pi _{os}=\rho _{0}k_{B}T\omega (c_{s}/\rho _{0})$ independent of any
model; the osmotic coefficient is also quite constant). At very high
concentrations, the Oosawa approximation does seem to underestimate
experimental pressures \cite{ref23}. This has been attributed to a failure
of the Oosawa model at close separations \cite{ref24} but a calculation not
invoking cylindrical symmetry still needs to be presented. Under conditions
of tight packing, it is reasonable to neglect undulatory entropy. The bulk
free energy is now the work of compressing the lattice from a state of
infinite separation to the state at hand all the while keeping the chemical
potential of the reservoir electrolyte fixed. 
\begin{equation}
F_{int}\left[ \rho _{0}\right] =Lf_{int}+const.N=-3^{\frac{1}{2}%
}L\int_{\infty }^{H}dH^{\prime }H^{\prime }\pi _{os}(H^{\prime })+const. N
\label{eq9}
\end{equation}
\begin{eqnarray}
f_{int} &=&\left( \frac{k_{B}T}{2Q}\right) I(w)  \label{eq10} \\
I(w) &\equiv &\int_{w}^{\infty }dy\frac{h(y)}{y}  \nonumber \\
&=&-h(w)-\frac{1}{2}\ln \left[ \frac{(1+w^{2})^{\frac{1}{2}}-1}{(1+w^{2})^{%
\frac{1}{2}}+1}\right]  \label{eq11}
\end{eqnarray}
Hence, the chemical potential $\mu =\partial F_{int}\left[ \rho _{0}\right]
/\partial N$ of a DNA monomer at a fixed volume and buffer concentration may
be expressed by 
\begin{equation}
\mu =const.+\frac{Ak_{B}T}{4Q}\ln \left[ \frac{(1+w^{2})^{\frac{1}{2}}+1}{%
(1+w^{2})^{\frac{1}{2}}-1}\right]  \label{eq12}
\end{equation}
Note that the limiting forms of the logarithmic term are $-2\ln w$\ as $w\ll
1$\ and $2/w$\ as $w\gg 1$; still , in practice, the variable $w$ is often
of order unity.

We next apply eq.(\ref{eq12}) to the deformed DNA lattice inside the phage whose
density $\rho (\mathbf{r})=4Qc_{s}/Aw(\mathbf{r})$\ at equilibrium is given
by eq.(\ref{eq6}).
\begin{equation}
\frac{(1+w^{2}(\mathbf{r}))^{\frac{1}{2}}+1}{(1+w^{2}(\mathbf{r}))^{\frac{1}{%
2}}-1}=k\exp \left( -\frac{2PQ}{R_{c}^{2}(\mathbf{r})}\right) \equiv Z(%
\mathbf{r})  \label{eq13}
\end{equation}
where $k$\ is a constant. This is rewritten as 
\begin{equation}
\rho (\mathbf{r}) = \left( \frac{2Qc_{s}}{A}\right) \frac{Z(\mathbf{r})-1}
{Z^{\frac{1}{2}}(\mathbf{r})}
\label{eq14}
\end{equation}
implying that the dimensionless function $Z(\mathbf{r})$ must be larger than
unity otherwise a solution to eq.(\ref{eq13}) does not exist. Without knowing the
precise structure of the DNA globule inside the phage, we already conclude
from eq.(\ref{eq13}) that tightly wound regions of curvature radius $R_{c}\simeq 
\sqrt{2PQ}\simeq 8$ nm must be depleted of DNA (persistence length $P$ = 50
nm; Bjerrum length $Q$ = 0.71 nm). The tight bending of the DNA causes an
increase in the curvature stress which has to be relieved by lowering the
electrostatic stress (see fig. 1).

Let us now consider the profile about a cylindrical hole within a phage
filled with DNA of genomic size. Then, the DNA almost fills the compartment
of volume $V$\ completely so we have from eq.(\ref{eq5}) 
\begin{equation}
\frac{L}{A}=\int_{V}d\mathbf{r\ }\rho (\mathbf{r}) \simeq
\frac{2Qc_{s}V(k-1)}{Ak^{\frac{1}{2}}}
\label{eq15}
\end{equation}
i.e. 
\begin{equation}
X\equiv \frac{L}{2Qc_{s}V}\simeq \frac{k-1}{k^{\frac{1}{2}}}  \label{eq16}
\end{equation}
By way of example, let us reconsider the data on phage T7 of Cerritelli at
al \cite{ref5} which we previously analyzed in terms of a model of uniform
density \cite{ref2}. The NaCl concentration of the buffer was 0.1 M. The DNA
genome of length $L$\ is enclosed in a volume $V$ of about 77600 nm$^3$ which
almost equals the volume $V_{DNA}$\ of the DNA lattice itself. Eq.(\ref{eq16})
then yields $X=2.06$ and $k=6.08$. Accordingly, the radius $E$ of the
(presumably single) hole free from DNA is computed from eq.(\ref{eq13}): 
\begin{equation}
E^{2}=\frac{2PQ}{\ln k}  \label{eq17}
\end{equation}
yielding $E\simeq 6$ nm. This should be compared with a hole radius of
about 4 nm or possibly larger, based on the experimental spacing
($H=2.54$ nm) and estimates of the inner volume of bacteriophage T7
\cite{ref2, ref5}. An alternative route to a theoretical estimate for
$E$ is via the (average) experimental spacing $\bar{H}$: the volume
$V_{DNA}$\ of the DNA lattice is slightly less than $V$\ and so
$V\simeq L \bar{S}$. Hence, we have $X=2.10$ from eq.(\ref{eq16}) and
again $E\simeq 6$ nm. Similar conclusions apply to recent structural
work on bacteriophage T4 \cite{ref25}. If the phage is modeled as an
ideal sphere of radius $R=41.5$ nm, the radius of the hole would be
about 10 nm (setting $L=57800$ nm; $c_{s}\simeq 0.3$ M) from
eq.(\ref{eq17}). Olson et al \cite{ref25} studied the DNA density which
decreases inward starting from the outer capsid shell and there does
appear to be a hole of radius $\sim 10$ nm. However, at present, it is
difficult to make this comparison more quantitative. Finally, we can
justify the neglect of density gradients and Coulomb integral terms in
eq.(\ref{eq2}); they are only of relative magnitude $H^{2}/R_{c}^{2}$
which is much smaller than unity.

We conclude from a simple model of close-packed DNA enclosed in phages, in
which we relax the supposition of uniformity, that a hole (or holes) must
occur free from DNA, at least at ionic strengths typical of laboratory or
physiological conditions. The curvature stress in a tightly wound region of
DNA becomes so high that the nonuniform Donnan effect can only compete up to
a point; the DNA is thus forced out of such a region beyond a certain
curvature given by eq.(\ref{eq17}) (See fig. 2).
It would be interesting to see
if the DNA density may be approximated by eq.(\ref{eq13}) -- which is independent
of the shape of the capsid (!) -- but one has to be wary
about the limitations of the present continuum analysis. At low densities,
toward a hole, the DNA mesophase should become more disordered. Kasspidou
and Van der Maarel have established that a hexagonal phase of short DNA rods
melts into a cholesteric phase \cite{ref9}. This happens, for instance, at an
interaxial spacing of $H=3.7$ nm in a 0.10 M NaCl solution. It is then
predicted from eq.(\ref{eq14}) that such a transition would occur at a radius of
curvature $R_{c} = 9$ nm in the T7 phage \cite{ref5}, at least if we are
allowed to extrapolate the results for short rods \cite{ref9} to long DNA.
At such nanoscales the concept of phases becomes very tentative of course.
Anyway, the more disordered cholesteric ''phase'' surrounding the hole would
still have an average spacing $H$ very similar to that of the coexisting
hexagonal phase \cite{ref9} in the bulk of the phage. Another possibility
for disorder is an escape of a section of the DNA ''into the third
dimension'' i.e. if a DNA strand aligns along the cylindrical axis of the
hole. Such a state would seem to be unstable however.

The buffers used in \cite{ref6, ref25} also contain magnesium ions though
not in excess. When genomic DNA is then packaged in bacteriophage T7, two
Raman bands of the DNA are significantly perturbed with respect to DNA in
solution \cite{ref26}. Overman et al attribute this to enhanced magnesium
concentrations within the phage \cite{ref26}. The effect of divalent ions
has not been allowed for here although we note that magnesium is absent in
the buffers used by Cerritelli et al \cite{ref5}.

Small angle X-ray scattering could be useful in trying to verify the profile
predicted by eq.(\ref{eq14}). However, it is well known that the scattering by
phages is subtle with regard to interpretation \cite{ref27}. The wavevector
dependence of the scattering by lyotropic liquid crystals is peculiar in
its dependence on the osmotic compressibility and the elastic moduli \cite
{ref28}. It might be useful to extend theories pertaining to scattering from
the bulk \cite{ref15, ref16, ref18} to that from the DNA mesophase confined
in phages. We remark that there is a length scale in the theory for a
fluctuating splayless liquid crystal which turns out to be identical to our
decay length $\sqrt{2PQ}$ in eq.(\ref{eq13}) (in \cite{ref15, ref16, ref18}, there
is also a competition of bending and osmotic stresses). Therefore, the DNA
liquid crystal near the outer rim of the phage behaves much like a useful
polymer liquid crystal but this is no longer true near the hole.

\bigskip

\section*{Acknowledgment}

We thank Paul Jardine for many discussions on viruses and Rudi Podgornik for
clarifying the status of the possible hexatic phase. One of us (F.S.) is
grateful to the Dutch Foundation for Virological Research for a grant
entitled "Virussen in katachtigen".

\newpage

\begin{figure}
  \centering
   \includegraphics[width=100mm]{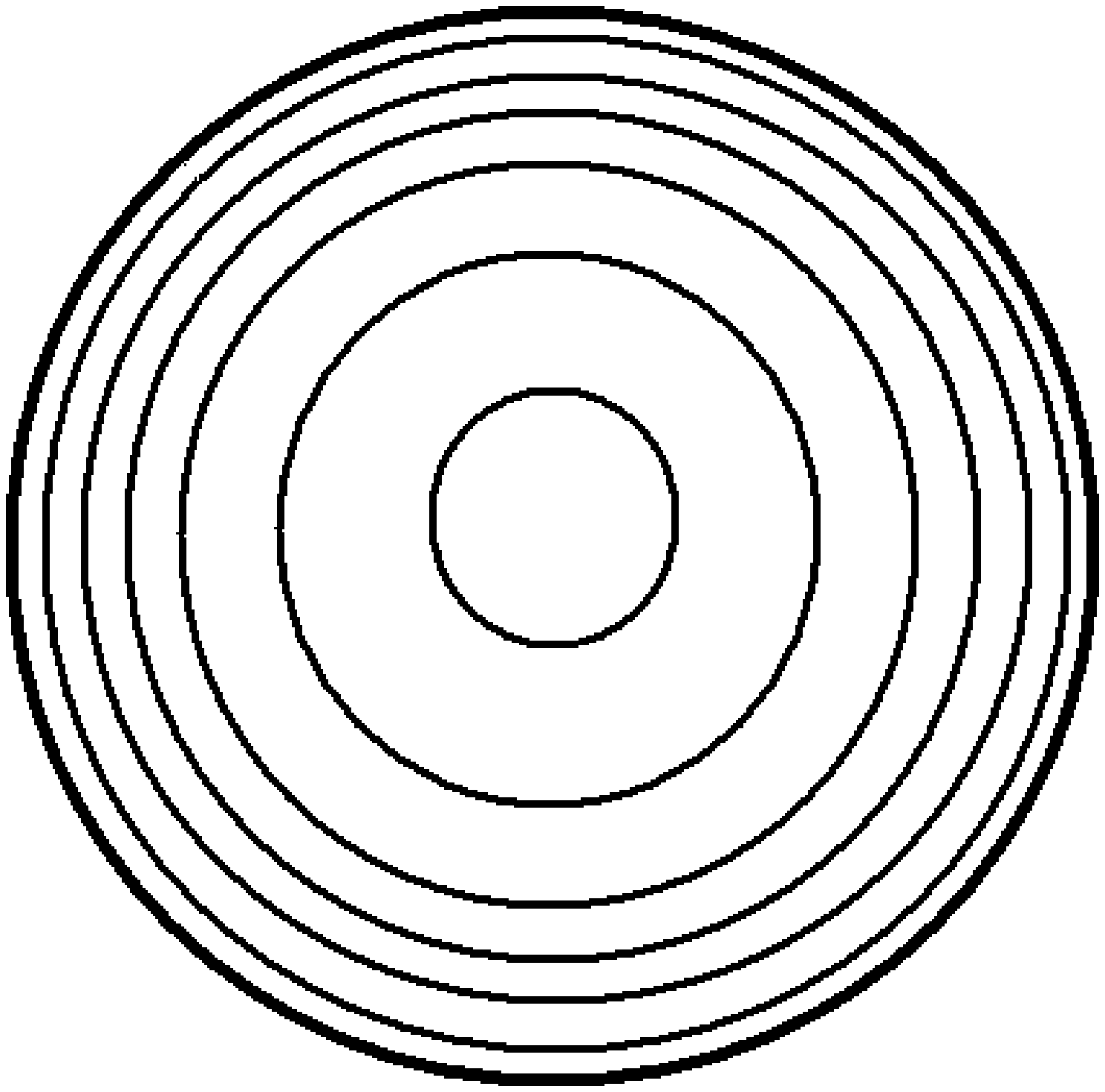}
   \caption{Schematic representation of DNA packed in a spherically
shaped virus (plane perpendicular to the spool axis). The packing is
locally hexagonal and becomes less close in the inner region of tight
winding.  }
\end{figure}

\begin{figure}
  \centering
   \includegraphics[width=100mm]{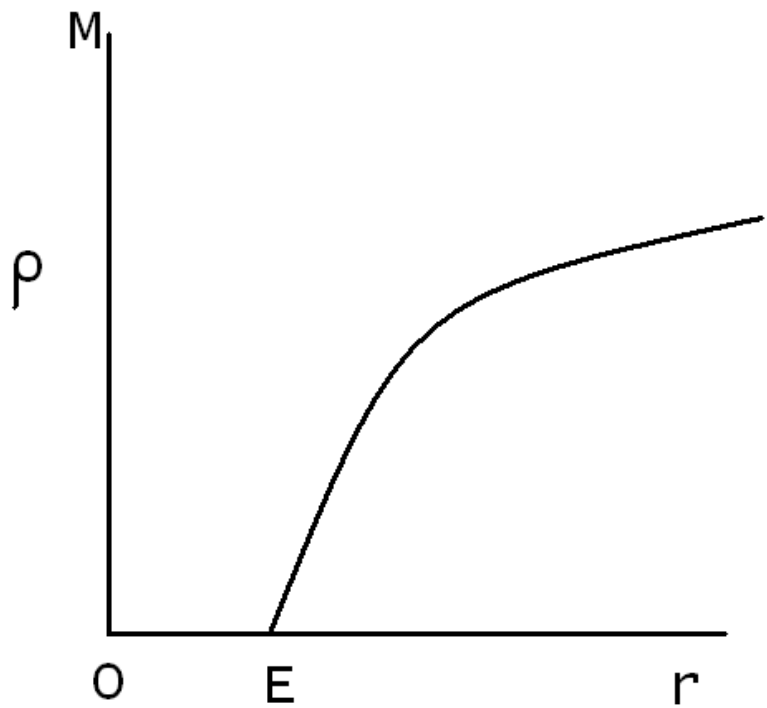}
   \caption{ Coarse-grained DNA density in a plane through the central
axis (MO) of a hole within a presumably spoollike region of DNA inside
the phage. The size of the nonuniform region beyond $E$ is
approximately equal to $\sqrt{2PQ}$. }
\end{figure}

\end{document}